\documentclass[conference]{IEEEtran}
\IEEEoverridecommandlockouts
\usepackage{cite}
\usepackage{amsmath,amssymb,amsfonts}
\usepackage{algorithmic}
\usepackage{graphicx}
\usepackage{textcomp}
\usepackage{xcolor}
\usepackage{float}
\usepackage{comment}
\def\BibTeX{{\rm B\kern-.05em{\sc i\kern-.025em b}\kern-.08em
    T\kern-.1667em\lower.7ex\hbox{E}\kern-.125emX}}
\begin{document}

\title{From Real-World Projects to Research-Oriented Learning: Continuous Improvement of a Master-Level Course in Software Engineering Education}

\author{\IEEEauthorblockN{1\textsuperscript{st} Michael Neumann}
\IEEEauthorblockA{\textit{Dpt. of Business Information Systems} \\
\textit{University of Applied Sciences and Arts Hannover}\\
Hannover, Germany \\
0000-0002-4220-9641}
\and
\IEEEauthorblockN{2\textsuperscript{nd} Eva-Maria Schön}
\IEEEauthorblockA{\textit{University of Applied Sciences Emden/Leer}\\
Emden/Leer, Germany \\
0000-0002-0410-9308}
}


\maketitle

\begin{abstract}
\textit{Problem:} Despite growing interest in project-based learning, little is known about how a master-level course can be continuously evolved toward research-oriented approaches over several years and how students perceive this development.
\textit{Method:} We conducted a longitudinal mixed-methods study of a master-level course in Information Systems at the University of Applied Sciences and Arts Hannover (Germany). The analysis covers six years between 2019 and 2025 and draws on teaching evaluations, course documentation, and reflective teaching artifacts.
\textit{Results:} The course evolved from a practice-oriented project format toward a more explicitly research-oriented learning environment. Despite this change, students' perceived course quality remained positive. Authentic projects, external collaboration, lecturer support, structured scaffolding, and visible relevance supported positive student perceptions.
\textit{Contribution:} This paper shows how a master-level course can be continuously evolved toward research-oriented learning while maintaining positive student perceptions. It further identifies the course design decisions that supported this transition.
\end{abstract}


\section{Introduction}
Research-oriented teaching formats play a pivotal role in software engineering education~\cite{Morais.2021}. They allow lecturers to combine methodological learning with practical problems, collaboration with external stakeholders, and tasks that reflect later professional or academic work~\cite{Fioravanti.2018}. Our prior work showed that students value such settings when they integrate real-world problems, research-related activities, and a high degree of self-organization and responsibility~\cite{BlindedSource1} .

However, such course formats are demanding to design and to teach~\cite{Perez.2020}. Once lecturers move beyond introductory project work and ask students to work on empirically grounded questions, justify methodological decisions, and produce results that meet higher expectations regarding rigor and quality. Students no longer only apply known concepts to a practical case~\cite{Jaiswal.2025}. They have to deal with uncertainty, make research decisions, structure their work more independently, and deliver outcomes that are expected to be both practically relevant and scientifically sound~\cite{Gupta.2022}.

This development is highly relevant in courses that are continuously evolved over several years. In such settings, changes are usually not limited to single teaching methods or organizational details~\cite{Faizi.2021}. Instead, lecturers often raise the demands step by step~\cite{Silva.2021}, for example by increasing methodological rigor, expanding the complexity of the projects, intensifying collaboration with practice partners, or integrating new forms of research support such as AI-based tools~\cite{Robol.2025}. From a teaching perspective, this evolution is desirable because it allows the course to reflect current research practice and professional requirements more closely. From the students’ perspective, however, increasing research complexity may also lead to higher workload, more cognitive demands, and more uncertainty in the learning process.

This development creates a central challenge for lecturers designing master-level courses in software engineering education. On the one hand, such courses should prepare students for complex professional and academic settings by integrating authentic projects, empirical inquiry, and increasing levels of autonomy~\cite{Prince.2025}. On the other hand, evolving a course toward stronger research orientation also introduces additional demands regarding methodological rigor, coordination effort, and support needs. The challenge is therefore not only whether students continue to evaluate such a course positively, but also how a course can be continuously refined so that research-oriented learning remains both feasible and valuable from the students' perspective.

Existing work on project-based and externally connected learning provides important foundations for this discussion~\cite{CehVarela.2023}. However, prior studies mostly focus on the implementation or perception of a particular pedagogical format at a given point in time (\textit{e.g.}, \cite{Neumann.2021}). They offer less insight into how a master-level course evolves over several years toward a research-oriented learning environment and which course design decisions support this evolution. In particular, there is still limited empirical knowledge on how authentic industry collaboration, methodological refinement, and structured support interact over time in a course that becomes increasingly oriented toward empirical research. Also the importance to collaborate with companies from different industries and public administrations and to integrate practice-relevant phenomenons into the course is underexplored.



Based on this motivation, we address the following research questions:\\
\textbf{RQ1:} How did the increasing research orientation of the course relate to students’ perceived course quality over time?\\
\textbf{RQ2:} How do the students value the research-oriented course design?\\
\textbf{RQ3:} Which elements of the evolving course design help maintain or improve course quality from the students’ perspective?
The paper at hand is structured as follows: We provide an overview of the related work in Section~\ref{sec2:RelWork}. Next, we explain our research approach in Section~\ref{sec3:ResearchApproach}. We present the results and answer our research questions in Section~\ref{sec4:Results} and discuss the study results to provide implications for other lecturers in Section~\ref{sec5:Discussion}. Finally, the paper closes with a conclusion and an outlook of the future work in Section~\ref{sec6:Conclusion}.

\section{Related Work}
\label{sec2:RelWork}
Active learning formats such as Problem-Based Learning (PBL), Project-Based Learning (PjBL), and Challenge-Based Learning (CBL) are widely used in higher education to foster participation, practical relevance, and competence development~\cite{Varela.2025}. Serrano Segarra~\cite{Segarra.2020} emphasizes that PBL supports professional competence development by shifting students from passive reception to active problem solving. Carella and Colombo~\cite{Carella.2024} similarly describe project-based learning as a way to strengthen experimentation, engagement, and the transfer of disciplinary concepts through collaboration between a university course and a company.

In software engineering education, project-based learning has also been reported as a promising way to connect theory and practice. Souza et al.~\cite{Souza.2019} show that students perceived project-based learning positively, especially with regard to teamwork, realistic development settings, and practical assignments. Related work further highlights the relevance of authentic learning environments and external partnerships. Netreba et al.~\cite{Netreba.2025}, for example, argue that professional-academic partnerships can create more meaningful and workplace-relevant learning experiences.

A broader conceptual perspective is provided by McGill et al.~\cite{McGill.2012}, who describe a teaching-research-industry-learning nexus in ICT education. Their work suggests that teaching, learning, research, and industry collaboration should be understood as interconnected dimensions of educational design. Gallagher and Savage~\cite{Gallagher.2020} similarly show that challenge-based learning can foster transversal competences and authentic collaboration, but they also point to a major limitation in the field: the heterogeneity of implementations makes it difficult to compare results across studies.

Research-oriented teaching has also been addressed more explicitly in higher education research. B{\"o}ttcher and Thiel~\cite{Boettcher.2018} present an instrument to assess students' research competences, including methodological skills, reflection on findings, and communication. Their work shows that research-oriented teaching can be linked to concrete competence dimensions, but it does not address how students perceive an evolving research-oriented course design over time.

However, the existing literature provides strong support for active, authentic, and externally connected learning formats. Two limitations remain. First, most studies focus on the implementation or perception of a specific pedagogical format rather than on the longitudinal evolution of one course over several years. Second, prior work commonly reports benefits of authenticity, collaboration, and project-based learning, but offers less insight into how students perceive a course when methodological rigor, research autonomy, and project complexity are systematically increased over time. In particular, there is still limited empirical knowledge on how a master-level course can be continuously evolved from practice-oriented project work toward a research-oriented learning environment, which design decisions support this evolution, and whether positive student perceptions can be maintained throughout this process.

\section{Research Approach}
\label{sec3:ResearchApproach}
This study follows a longitudinal mixed-methods research approach. We analyze six consecutive years of teaching evaluations from a research-oriented master-level course in Information Systems and complement the quantitative data with course documentation and reflective teaching artifacts. In this way, we trace how the course evolved between 2019 and 2025 and how increasing research complexity relates to course quality from the students' perspective.

\subsection{Research Context}

\paragraph{Study Program}
The study was conducted in a master-level course at the University of Applied Sciences and Arts Hannover, Germany. More specifically, the course is part of the Master of Science program \textit{Digital Transformation}, which has been offered by the Business Computing department since the winter term 2018/2019. A typical student cohort in the program comprises about 25 students, and the course is offered annually in the summer term. 

\paragraph{Course Information}
The course was originally designed to familiarize students with innovative methods and current challenges of project management and to enable them to address these challenges using systematic approaches and methods. The learning objectives include understanding agile, plan-based, and hybrid approaches, dealing with intercultural and distributed teams, leadership and team coordination, conflict management, and the presentation of results for selected stakeholders. From the beginning, the course therefore combined methodological learning with project-based work and stakeholder-oriented communication. The total student workload was planned as 180 hours, including 68 hours of in-class teaching and 112 hours of self-study.

The learning objectives of the course focused on enabling students to understand and apply central methods and challenges of project management in a systematic way. In particular, the course addressed agile, plan-based, and hybrid approaches, the management of intercultural and distributed teams, leadership and team coordination, conflict management, stakeholder-oriented reporting, and the personnel management and motivation of virtual international teams.

The form of examination is a seminar paper with two components: a written paper and a presentation. This examination format reflected the general course design, as students were expected not only to work on a practically relevant problem during the term, but also to document and present their work in a structured and academically sound way. This basic orientation towards methodologically grounded, project-based, and stakeholder-related work remained central to the course and forms the basis for the later evolution towards a more explicitly research-oriented course design.

\paragraph{Course Design} A central design element of the course was the integration of real-world projects in collaboration with companies from different industries and public administrations. In the past we collaborated with companies, e.g., from the insurance, e-commerce, online marketing, or automotive sectors. The student cohort was split into four teams and each team worked with one associated company and one specific project. The selected projects addressed real problems from industries such as banking, insurance, energy, retail, or chemicals. In this setting, the course was not limited to discussing methods in an abstract way. Instead, students had to apply methods to authentic project contexts, interact with external stakeholders, and work towards results that were relevant for the respective partner organization. 

To organize this work, we applied an adapted eduScrum approach~\cite{BlindedSource1}. Since the original eduScrum framework was developed for school education~\cite{Wijnands.2019} and not for higher education settings with real-world stakeholders, we adapted its roles, events, and artifacts to our needs. In particular, the lecturer took the role of an \textit{Agile Coach}, external practitioners acted as \textit{Product Owners}, each student team selected an \textit{eduScrum Master}, and the student teams worked as self-organizing groups. The Product Owners were responsible for the requirements and the product backlog, while the Agile Coach supported the students with methodological, organizational, and content-related questions. This design allowed us to combine structured guidance with a high degree of student responsibility and autonomy. 

The adapted course design also included iterative work practices and concrete coordination artifacts. The student teams worked in two-week sprints and used meetings such as sprint planning, review, retrospective, and refinement. In addition, they used artifacts such as product backlogs, sprint backlogs, definitions of done, and product increments. These elements were introduced not only to structure the course process, but also to align the educational setting more closely with real project work in practice. The shorter sprint length was deliberately chosen to support tighter coordination between the student teams, the lecturers, and the external Product Owners.

\subsection{Data Collection}
The main data source of this study consists of teaching evaluation questionnaires from the course iterations in 2019, 2021, 2022, 2023, and 2025. No evaluation data were included for 2020 and 2024. We excluded 2020 because the Covid-19 pandemic led to an abrupt shift from on-site teaching to emergency remote teaching, which substantially changed the course setting and limited comparability with the regular course implementations. For 2024, no usable evaluation data were available because the number of returned questionnaires was too low to support a meaningful analysis.

Across the included years, the evaluation questionnaires were not fully identical, but they showed clear overlap in their central focus. The 2019 questionnaire followed a broader structure with separate sections on participation, workload, course content, lecturer evaluation, and didactics, and partly used different response formats, including six-point scales. In contrast, the questionnaires from 2021 onwards relied on more compact sets of closed agreement items on a five-point scale. Despite these differences, the questionnaires repeatedly addressed closely related aspects such as course structure, active participation, comprehensibility of instruction, teaching methods, learning atmosphere, and students' overall perceptions of the course.

A particularly stable core of questionnaire items can be observed from 2021 onwards. The questionnaires from 2021, 2022, and 2023 include highly similar items on the structure of the course, opportunities for active participation, the comprehensibility of the lecturer's explanations, variation in teaching methods, students' contribution to the success of the course, the working atmosphere, the social climate between students and lecturer, and the stimulation of independent thinking. In 2022 and 2023, this core was extended by additional global items on perceived learning gain, a learning-supportive atmosphere, and recommendation of the course. The questionnaire used in 2025 remained largely compatible at the construct level, although some wording was adapted and additional items were introduced.

In addition to the closed questions, the questionnaires also contained open-text fields on positive aspects of the course, criticism, and suggestions for improvement. Although the wording varied across years, these responses provided complementary qualitative insights into how students perceived the course design, workload, practical relevance, and research-related activities. We use these open responses as contextual material for interpreting the longitudinal evaluation results, while the main longitudinal comparison is based on the recurring dimensions of the closed items.

Table~\ref{tab:evaluation-comparability} summarizes the comparability of the evaluation instruments across the observed course iterations. While the questionnaires were not fully identical across all years, they showed clear overlap in central evaluation dimensions. In particular, the questionnaires from 2021 onwards provide a stable core for longitudinal comparison, whereas the 2019 questionnaire can be linked to the later instruments on the level of broader constructs rather than identical items.

\begin{table}[t]
\caption{Overview of questionnaire comparability across years}
\label{tab:evaluation-comparability}
\centering
\begin{tabular}{p{0.9cm}p{3.2cm}p{2.8cm}}
\hline
\textbf{Year} & \textbf{Main characteristics} & \textbf{Longitudinal use} \\
\hline
2019 & Different questionnaire structure with separate thematic blocks and partly different scales & Construct-level reference point \\
2021 & Compact 5-point questionnaire with recurring core items & Start of robust longitudinal comparison \\
2022 & Same core design as 2021, plus global evaluation items & Direct comparison possible \\
2023 & Nearly identical to 2022 & Direct comparison possible \\
2025 & Largely compatible core dimensions, but some adapted and added items & Partial comparison on recurring dimensions \\
\hline
\end{tabular}
\end{table}

The use of multiple data sources was important for two reasons. First, the teaching evaluations provide information on course quality from the students' perspective, which is central to our research questions. Second, the additional course-related artifacts (such as product backlogs or research protocols created by the students) allow us to analyze how research complexity changed across the observed years and which design decisions may have contributed to stable or positive student perceptions.

\subsection{Data Analysis} We analyzed the teaching evaluations with a longitudinal perspective and distinguished between two levels of comparability. First, we used the strongly overlapping questionnaire versions from 2021, 2022, 2023, and partly 2025 to compare recurring core dimensions over time. These dimensions include, among others, perceived course structure, active participation, comprehensibility of instruction, learning atmosphere, and the stimulation of independent thinking. This allows for a robust longitudinal comparison across the later course iterations.

Second, we included the 2019 questionnaire at a more abstract level of analysis. Although its exact item structure differs from the later instruments, it addresses related areas such as course content, didactics, lecturer quality, workload, and practical relevance. We therefore use 2019 as an earlier reference point on the level of broader evaluation dimensions rather than as a fully item-equivalent measurement point. This enables us to reconstruct the longer-term development of student perceptions without overstating the degree of measurement equivalence across all years.

Overall, this procedure results in a longitudinal mixed-methods analysis with a stable core of evaluation dimensions. It allows us to examine developments in students' perceived course quality over time while taking into account that the questionnaires evolved together with the course itself.

\subsection{Threats to Validity}
As with every study, our work is subject to certain limitations. To increase transparency, we discuss the main threats to validity following the framework by Runeson and Höst~\cite{Runeson.2009}.

\paragraph{Construct validity}
A first threat concerns the operationalization of the central concepts of this study, namely \textit{research complexity} and \textit{course quality from the students' perspective}. Course quality is mainly captured through teaching evaluations and thus through students' subjective perceptions, which do not represent all possible dimensions of course quality. Likewise, research complexity is not measured through a single variable, but reconstructed from the documented evolution of the course. To reduce this threat, we complemented the evaluation data with course documentation and reflective teaching artifacts.

\paragraph{Internal validity}
A second threat concerns the attribution of observed changes in student perceptions to the evolving course design. Since the course changed over several years, multiple factors may have influenced the results, including cohort effects, differences in prior experience, changes in project settings, or broader changes in the teaching context. We therefore do not derive strong causal claims, but interpret the evaluation results in relation to documented course changes.

\paragraph{External validity}
A third threat concerns the generalizability of the findings. The study investigates one research-oriented master-level course in one specific study program at one German University of Applied Sciences. Consequently, the findings cannot be transferred directly to other institutions or course settings without considering these contextual conditions. At the same time, the study provides a detailed longitudinal account that may inform similar courses in software engineering education.

\paragraph{Conclusion validity}
A final threat concerns the extent to which the data support the conclusions drawn in this study. While the longitudinal evaluation data and complementary qualitative artifacts provide a useful basis for interpretation, teaching evaluations are not designed to explain effects on their own. We therefore combine quantitative and qualitative data, focus on recurring patterns across course iterations, and formulate our conclusions cautiously.

\section{Results}
\label{sec4:Results}
In this section, we first provide an overview of the teaching evaluations results. We then describe the longitudinal evolution of the course design from practice-oriented project work toward a more explicitly research-oriented learning environment. Based on this course evolution, we analyze how students perceived the course over time and which elements of the evolving course design supported positive student perceptions.

\begin{figure*}
\centering
	\includegraphics[scale=0.214]{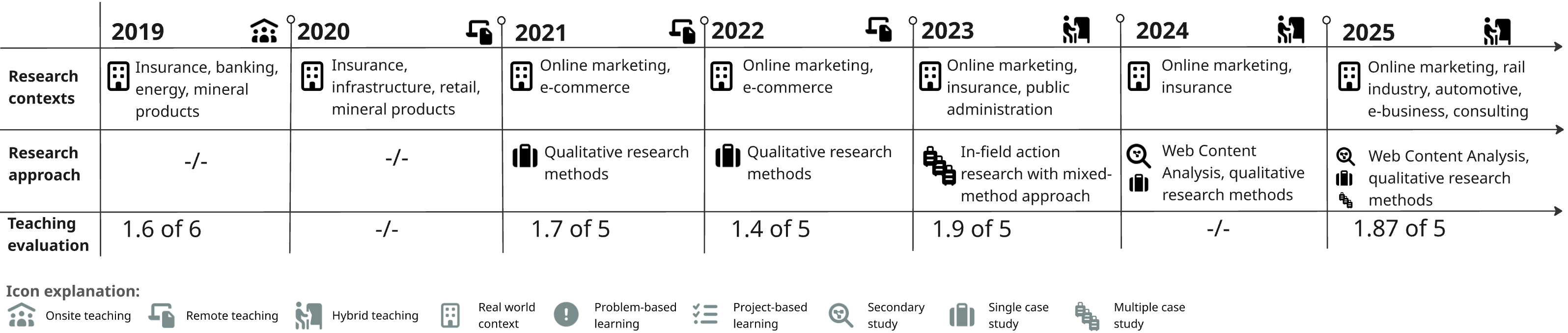}
\caption{Timeline of the course evolution}
\label{fig:timeline}
\end{figure*}

\subsection{Overview of the Teaching Evaluation Results}
Across the available years, the results show a visible decline in the number of returned questionnaires. While the early course iterations yielded double-digit numbers of responses, the 2025 evaluation included only seven returned questionnaires. This reduction limits the robustness of year-to-year comparisons and therefore needs to be considered when interpreting longitudinal trends.

In addition, the questionnaires show some item-level non-response, especially in the 2019 evaluation. Overall, however, the available dataset provides a sufficient basis for descriptive longitudinal comparisons on the level of recurring evaluation dimensions, while conclusions for single years with smaller numbers of returned questionnaires must be interpreted with appropriate caution.

\subsection{Longitudinal Evolution of the Course Design}
The course evolved substantially between 2019 and 2025 (see Figure~\ref{fig:timeline}). In the early course iterations, it was primarily characterized by authentic industry projects and practice-oriented analyses. At this stage, the course already emphasized real-world relevance and external collaboration, but the projects were not yet consistently framed as explicit empirical research studies. 

From 2021 onwards, the course moved more clearly toward research-oriented learning. Additional thematic focal areas were introduced, including topics such as remote work, workload, autonomy, and agile teamwork. At the same time, the empirical character of the projects became more explicit, for example through qualitative case studies and a stronger emphasis on research questions, data collection, and methodological justification.

In 2022, this design remained largely stable, while 2023 marked a further step in the course evolution. The projects expanded beyond single case studies and included more advanced designs such as multi-case studies and action research, especially in the context of experiments on compressed work schedules. Here, multiple student groups investigated the same topic using shared research questions in different field contexts. In the action research projects, this also required the groups to coordinate and orchestrate the interventions across contexts, which led to considerably higher inter-group alignment efforts.

In 2024 and 2025, the methodological spectrum broadened further. The projects included conceptual studies, secondary studies, workshop-based and mixed approaches, as well as multi-case designs. At the same time, the thematic focus shifted towards more complex and novel topics such as agile maturity models, Objectives and Key Results implementation, communication in transformation projects, agile coaching impact with machine learning, neurodiversity in agile teams, and GenAI adoption in agile software development.

Overall, the course evolved from a practice-oriented project format into a more explicitly research-oriented learning environment. This evolution did not only increase methodological demands, but also changed the role of the students, the role of the lecturer, and the forms of coordination required within and across projects.

\subsection{Student Perceptions Over Time}
Here, we answer RQ1: \textit{How did the increasing research orientation of the course relate to students’ perceived course quality over time?}

To analyze student perceptions over time, we first considered the global indicators reported in the evaluation summaries for each available course iteration. Across all five included years, the course was evaluated positively. Although some variation can be observed between individual years, the overall pattern remains stable and does not indicate a substantial decline in students' perceived course quality.

This finding is particularly noteworthy in light of the longitudinal evolution of the course design. During the observed period, the course developed toward a more explicitly research-oriented learning environment with broader methodological repertoires, more demanding empirical work, and increasingly complex thematic and organizational settings. Despite this development, the evaluation results do not suggest that students' perceptions deteriorated in a substantial way.

A similar pattern emerges when focusing on recurring evaluation dimensions from the more comparable questionnaire versions from 2021 onwards. Across these years, the indicators related to course structure, active participation, comprehensibility of instruction, social climate, and independent thinking remained positive overall. At the same time, the qualitative responses indicate that this stability should not be interpreted as the absence of tension. Students repeatedly referred to higher workload, ambiguity, effort for data collection and analysis, and coordination challenges in group work. Thus, the results do not suggest that increasing research orientation is unproblematic. Rather, they indicate that growing methodological and organizational demands can remain compatible with positive student perceptions when the course design provides sufficient orientation, support, and visible relevance.

Overall, the findings suggest that the relationship between research-oriented course evolution and perceived course quality is not best understood as a simple trade-off. Instead, stronger research orientation did not lead to a general decline in perceived course quality, provided that increasing demands were accompanied by appropriate scaffolding and support.

\subsection{Student Perceptions of the Research-Oriented Course Design}
In this subsection we answer RQ2: \textit{How do the students value the research-oriented
course design?}

The open responses show that students generally valued the research-oriented course design positively, especially when it combined authentic projects, collaboration with companies, and a high degree of autonomy. Across multiple course iterations, students explicitly appreciated the work with real organizations, the practical relevance of the projects, and the opportunity to engage in a course format that differed from more traditional lecture-based teaching. In 2021, for example, students highlighted ``\textit{real companies}'', ``\textit{project work}'', ``\textit{a lot of freedom}'', ``\textit{practical experience}'', and a teaching method that was clearly different from usual courses. Similar statements can be found in 2022, where students positively referred to the company partners, the project work, the modern approach of the module, the possibility to work on the examination during the semester, and the support within the project. In 2023 and 2025, students again emphasized practice orientation, collaboration with external companies, interesting research partners, and the fact that the projects were perceived as something new and meaningful.

A second recurring theme in the open responses is the positive valuation of autonomy and active research work. In 2021, students explicitly valued ``\textit{self-organization}'', ``\textit{self-responsibility}'', and ``\textit{independent work}''. In 2023, they positively referred to the self-directed conduct of interviews and questionnaires, the opportunity to collect data themselves, and the ``\textit{freedom in research}'' that allowed them to work creatively. In 2025, students again highlighted self-organization, coaching, and the value of the course as preparation for the master's thesis. Taken together, these responses indicate that students did not merely tolerate the research-oriented design, but often perceived its central characteristics as beneficial and appropriate for a master's-level course.

At the same time, the open responses also show clear tensions in how students valued this course design. Several students criticized the high workload, the scope of the project work, the effort for data collection and analysis, and the combination of different assessment elements. In 2021, students mentioned time pressure, broad project scope, and an initially unclear thematic framing. In 2022, criticism focused on unclear research questions, group assignment, and the dependence of individual performance on team dynamics. In 2023 and 2025, students again pointed to stress in group work, extensive workload, the effort for interviews, questionnaires, and data analysis, and the perceived mismatch between the module title and the actual focus on research-related work. These responses suggest that students valued the research-oriented course design particularly for its authenticity, autonomy, and practical relevance, but also experienced its demands, ambiguity, and collaborative dependencies as substantial challenges.

\subsection{Course Design Elements Supporting Positive Student Perceptions}
Here we answer our RQ3: \textit{Which elements of the evolving course design help maintain or improve course quality from the students’ perspective?}

The qualitative responses suggest that positive student perceptions were supported by a combination of recurring course design elements rather than by one isolated intervention. Across the available evaluation waves, students repeatedly valued the practical and authentic character of the course. In 2022, students positively referred to the ``\textit{collaboration with companies}'' and the ``\textit{modern approach of the module}.'' In 2023, they described the ``\textit{collaboration with an external company}'' as ``\textit{really instructive and exciting}'' and explicitly emphasized the ``\textit{high degree of practical experience}.'' In 2025, students again highlighted ``\textit{interesting research partners},'' ``\textit{practical relevance},'' and stated that ``\textit{the collaboration with a company is very exciting}.'' These responses indicate that visible real-world relevance and collaboration with external partners were central elements supporting positive course quality from the students' perspective.

A second recurring element concerns the support structure provided by the lecturer. Students repeatedly valued direct support, accessibility, and opportunities for regular exchange. In 2022, they referred to ``\textit{support within the project},'' ``\textit{a great deal of support},'' and a lecturer who was ``\textit{reachable at any time and at very short notice}.'' In 2025, students mentioned ``\textit{weekly meetings},'' ``\textit{genuine support from the lecturer},'' and described the ``\textit{coaching}'' as ``\textit{very good for simply getting into an exchange}.'' The closed items from the same years support this pattern: accessibility of the lecturer was rated very positively in 2025 (mw = 1.29), and comprehensible explanations and a positive climate between students and lecturer were also rated strongly in 2022 and 2023. This suggests that high autonomy and demanding project work were perceived positively especially when they were accompanied by visible lecturer support. 

The evaluations further show that students valued elements that structured and scaffolded the research-oriented work. In 2023, students positively mentioned a ``\textit{well-structured course}'' and found the ``\textit{workshop on scientific work very helpful}.'' In 2025, they referred to the ``\textit{structure of the sessions}'' and explicitly appreciated the combination of ``\textit{a high degree of self-organization with initial recaps on scientific work}.'' These comments are important because they show that students did not simply value freedom as such. Rather, they valued a course design that combined self-organization with orientation, structure, and targeted support for scientific work.

At the same time, the negative comments clarify which elements became critical when this balance was weakened. In 2021, students reported that the ``\textit{initial orientation phase at the beginning of the project/course was difficult}'' and suggested ``\textit{one or two lectures at the beginning on the scientific framework}.'' In 2022, one student asked: `\textit{`What exactly are the research questions?}'' In 2023, students criticized the ``\textit{too much effort required for data collection and analysis},'' and in 2025 they pointed to a ``\textit{very high time commitment}'' and a ``\textit{high effort for data collection}.'' Taken together, these responses suggest that course quality was maintained when authenticity, external collaboration, and self-organization were combined with strong support, clear structure, and sufficient scaffolding of research activities.

\section{Discussion}
\label{sec5:Discussion}
The results of this study show that the longitudinal evolution of the course toward research-oriented learning did not lead to a general decline in perceived course quality. Although the course became more methodologically demanding, included broader empirical designs, and required more coordination and autonomy from students, the evaluation results remained positive over time. This finding complements prior work showing that students value project-based and authentic learning formats in software engineering and higher education~\cite{Souza.2019,Segarra.2020,Carella.2024}. At the same time, the present study extends this body of knowledge by indicating that such positive student perceptions can remain stable even when a course is continuously developed toward more explicit empirical research and stronger methodological demands. Rather than pointing to a simple negative effect of increasing rigor, the findings suggest a more differentiated picture in which growing demands can remain compatible with positive course evaluations when they are embedded in a meaningful and well-supported learning environment.

This finding also sharpens the interpretation of research-oriented teaching in software engineering education. Existing work has already shown that students appreciate project-based, authentic, and externally connected course formats~\cite{Souza.2019,Carella.2024,Netreba.2025}. The present results build on this perspective by showing that positive student perceptions can remain stable even when such a course is continuously evolved toward more explicit empirical research. In this sense, the paper adds a longitudinal perspective to a body of knowledge that has so far mainly examined individual implementations or single course iterations. What appears to matter is not the absolute level of difficulty alone, but the way in which this evolution is designed. In our case, the course did not simply become more demanding. It was continuously refined through authentic projects, collaboration with external partners, structured support, and increasing methodological guidance. This observation is consistent with the teaching-research-industry-learning nexus described by McGill et al.~\cite{McGill.2012}, while extending it by showing how such a nexus can be developed and sustained across multiple years of course evolution.

The qualitative responses nevertheless make clear that this development was not without tension. Students repeatedly referred to higher workload, ambiguity, effort for data collection and analysis, and coordination challenges in group work. These tensions matter because they show that research-oriented learning should not be romanticized. In this respect, the findings complement broader discussions on authentic and challenge-based learning, which likewise point to the heterogeneity and implementation complexity of such formats~\cite{Gallagher.2020}. Stronger research orientation should therefore not be understood as unproblematic by default. The more plausible conclusion is that methodological and organizational demands can remain compatible with positive student perceptions when the course design provides sufficient orientation, lecturer support, and visible relevance of the students' work.

Viewed more broadly, the relationship between research-oriented course evolution and perceived course quality is not best understood as a simple trade-off. Prior work on research-oriented teaching has primarily emphasized competence-related dimensions such as methodological skills, reflection, and communication~\cite{Boettcher.2018}. The present study adds that the longitudinal design of the course environment also matters for how such teaching is perceived by students. In our case, the course remained positively perceived because increasing autonomy and methodological demands were accompanied by recurring design elements that made the learning setting manageable and meaningful from the students' perspective.

\subsection{Implications for Lecturers}
For lecturers, the results suggest that research-oriented teaching can be designed in a way that maintains positive course quality even when methodological demands increase. This observation extends prior work on project-based, authentic, and industry-connected learning by translating it into a longitudinal course design perspective~\cite{Carella.2024,Netreba.2025,McGill.2012}. At the same time, the findings indicate that such outcomes do not emerge automatically from authentic or research-related tasks alone, but require a deliberate design of the course environment. Four implications appear particularly important.

First, the practical and research relevance of the course should be made highly visible. Across multiple years, students valued the authenticity of the projects, the collaboration with external partners, and the opportunity to work on problems that were perceived as meaningful beyond the classroom. This is consistent with prior studies on project-based learning and professional-academic partnerships, which emphasize the motivating role of authentic contexts and external stakeholders~\cite{Souza.2019,Carella.2024,Netreba.2025}. Our results add that this relevance becomes especially important when a course evolves toward more research-oriented learning and students are expected to invest substantial effort in empirical work.

Second, stronger research orientation requires explicit scaffolding. Students appreciated autonomy and self-organization, but they also demanded clearer framing of research questions, stronger orientation at the beginning of the course, and targeted support for research-related activities. In this respect, the findings complement research on research-oriented teaching that has so far focused more strongly on competence development than on course design conditions~\cite{Boettcher.2018}. Lecturers should therefore not only increase methodological rigor, but also invest in support structures such as introductory sessions, intermediate feedback, recaps on research work, and the explicit discussion of methodological decisions.

Third, lecturer accessibility remains central in demanding course formats. The qualitative responses show that students highly valued quick feedback, direct support, regular meetings, and a supportive exchange with the lecturer. Increasing autonomy therefore does not reduce the need for teaching presence. On the contrary, research-oriented course designs seem to require a strong instructional and coaching role in order to help students deal with uncertainty and complexity. This observation further specifies the teaching-research-industry-learning nexus described by McGill et al.~\cite{McGill.2012} by highlighting the lecturer's role as an active mediator between authentic projects, empirical inquiry, and student learning.

\begin{table*}[t]
\caption{Publication-oriented outcomes emerging from the course across different iterations}
\label{tab:publication-outcomes}
\centering
\begin{tabular}{p{0.8cm}p{2.2cm}p{2.4cm}p{5.6cm}p{3cm}p{2.2cm}}
\hline
\textbf{Year} & \textbf{Industry} & \textbf{Study Design} & \textbf{Topic} & \textbf{Venue} & \textbf{Status} \\
\hline
2021 & Online Marketing & Single case study & How a 4-Day Work Week and Remote Work Affect Agile Software Development Teams~\cite{Topp.2022} & Intl. Conf. on Lean and Agile Software Development & Published \\
2021 & E-Commerce & Qualitative single case study & Context factors of good remote work~\cite{Rometsch.2022} & Wirtschaftsinformatik \& Management & Published \\
2022 & E-Commerce & Qualitative single case study & Impact of Remote Work on Interaction and Autonomy in Agile Teams~\cite{Gill.2022} & Projektmanagement \& Vorgehensmodelle & Published \\
2022 & E-Commerce & Single case study & Workload effects and software development team outcomes~\cite{Sanden.2022} & HMD Praxis der Wirtschaftsinformatik & Published \\
2022 & Online Marketing & Single case study & Hybrid work organization in agile software development teams~\cite{Neumann.2022a} & Intl. Conf. in Software Engineering Research and Innovation & Published \\
2023 & Online Marketing & Single case study & Cultural diversity in agile software development teams~\cite{Welsch.2024} & Proc. of the ACM/SIGAPP Symposium on Applied Computing & Published \\
2023 & Public Administration and insurance & Multi-case study / action research & Compressed work schedules effects on stress and performance~\cite{Neumann.2025} & Euromicro Conf. on Software Engineering and Advanced Applications & Published \\
2024 & Online Marketing & Secondary / conceptual study & Impact of agile coaching on team performance & -/- & Under revision\\
2024 & Insurance & WCA \& Single case study & OKR framework in practice~\cite{Misslisch.2025} & Euromicro Conf. on Software Engineering and Advanced Applications & Published\\
2025 & Railway & Single case study & Neurodiversity in agile teams & -/- & Under revision\\
2025 & Automotive, consulting, ebusiness & Multi-case study & GenAI adoption pattern in agile software development teams~\cite{Neumann.2026} & Intl. Conf. on Agile Software Development & Accepted \\
\hline
\end{tabular}
\end{table*}

Fourth, workload and coordination effort should be treated as design issues rather than side effects. The open responses repeatedly show that students perceived data collection, analysis, group coordination, and the combination of different assessment elements as demanding. If research-oriented courses are intended to remain positively perceived, such demands have to be planned carefully. This includes realistic scoping of project work, transparent communication of expectations, and the alignment of research tasks with the available time and resources.

\subsection{Engaging Students for Science}
A central question behind this study is how students can be engaged not only in project work, but in scientific work. The findings suggest that this engagement does not primarily emerge from abstract methodological instruction alone. Students appear to engage with research when scientific work is connected to authentic contexts, relevant questions, and visible outcomes. In this sense, the course did not treat research as an isolated academic exercise, but as a meaningful way of understanding and addressing complex practice-oriented problems. This interpretation aligns with the broader literature on authentic, challenge-based, and externally connected learning~\cite{Gallagher.2020,Carella.2024,McGill.2012}, while extending it by showing how such engagement can be sustained in a longitudinally evolving master-level course.

A further point concerns the relationship between research ideals and student expectations. Students valued interviews, questionnaires, data collection, and the opportunity to work independently on their own ideas. Yet these same elements were also perceived as demanding, time-intensive, and at times unclear. Engagement for science cannot therefore be assumed simply because a course includes empirical methods or research tasks. It has to be actively fostered through course design. In this respect, the study complements prior work on research-oriented teaching that has mainly conceptualized such teaching through competence dimensions~\cite{Boettcher.2018} by showing that the surrounding course design conditions are equally important for how students experience this form of learning.

This dynamic is also reflected in the scientific output that emerged from the course over time. The publication overview shows that the course repeatedly led to publication-oriented outcomes across different years, research topics, and partner settings. This indicates that the course did not only motivate students to complete demanding projects, but in several cases enabled them to continue this work in the form of scientific publications or submissions. Under these conditions, research became visible not merely as a course requirement, but as a meaningful continuation of academic work. An overview of the publication outcomes is given in Table~\ref{tab:publication-outcomes}.

Publication-oriented outcomes should nevertheless not be treated as a mandatory success criterion. The open responses show that students perceived the research-oriented design as worthwhile, but also as effortful and ambiguous. Publication-oriented continuation cannot therefore be expected from all students equally. It is better understood as an indicator that the course was able to create a learning environment in which research became visible as a relevant and attractive form of academic practice.

The broader implication is that engaging students for science in software engineering education is possible without sacrificing perceived course quality. Such engagement does not emerge automatically through higher rigor alone. It depends on a course design that combines methodological demands with authenticity, autonomy with support, and scientific work with visible relevance. The contribution of this paper to the existing body of knowledge on project-based, authentic, and research-oriented learning lies in showing that engagement for science is not simply a consequence of adding empirical tasks, but of deliberately connecting research work to meaningful contexts, structured support, and visible outcomes. 

The results point to four design principles for research-oriented teaching in software engineering education: visible authenticity and external relevance, a gradual increase in research orientation, the combination of autonomy and scaffolding, and strong lecturer support and coordination. These principles capture the conditions under which the course could be continuously evolved toward research-oriented learning while maintaining positive student perceptions.

\section{Conclusion \& Future Work}
\label{sec6:Conclusion}
This paper showed how a master-level course can be continuously evolved toward research-oriented learning while maintaining positive student perceptions. Based on a longitudinal mixed-methods study, the results indicate that the course remained positively perceived even as it became methodologically broader, more explicitly research-oriented, and more demanding in terms of coordination and empirical work.

The findings further indicate that this stability depended on specific design elements, including authentic projects, collaboration with external partners, strong lecturer support, structured scaffolding, and visible relevance of the students' work. At the same time, the qualitative responses show that stronger research orientation also introduced tensions, especially regarding workload, ambiguity, and effort for data collection and analysis. Overall, the study contributes empirically grounded insights into how research-oriented teaching can be developed over time without losing positive student perceptions.

Future work should investigate similar course formats in other contexts, include additional indicators beyond teaching evaluations, and examine more closely the role of AI-supported research practices in research-oriented learning environments.


\bibliographystyle{IEEEtran}
\bibliography{references}

\end{document}